# Solid Source Metal-Organic Molecular Beam Epitaxy of Epitaxial RuO$_2$

William Nunn, Sreejith Nair, Hwanhui Yun, Anusha Kamath Manjeshwar, Anil Rajapitamahuni,

Dooyong Lee, K. Andre Mkhoyan, and Bharat Jalan

Department of Chemical Engineering and Materials Science

University of Minnesota,

Minneapolis, Minnesota 55455, USA

*Corresponding author: bjalan@umn.edu

A seemingly simple oxide with a rutile structure, $RuO_2$ has been shown to possess several intriguing properties ranging from strain-stabilized superconductivity to a strong catalytic activity. Much interest has arisen surrounding the controlled synthesis of $RuO_2$ films but, unfortunately, utilizing atomically-controlled deposition techniques like molecular beam epitaxy (MBE) has been difficult due to the ultra-low vapor pressure and low oxidation potential of Ru. Here, we demonstrate the growth of epitaxial, single-crystalline $RuO_2$ films on different substrate orientations using the novel solid-source metal-organic (MO) MBE. This approach circumvents these issues by supplying Ru using a "pre-oxidized" solid metal-organic precursor containing Ru. High-quality epitaxial $RuO_2$ films with bulk-like room-temperature resistivity of 55 $\mu\Omega$-cm were obtained at a substrate temperature as low as 300°C. By combining X-ray diffraction, transmission electron microscopy, and electrical measurements, we discuss the effect of substrate temperature, orientation, film thickness, and strain on the structure and electrical properties of these films. Our results illustrating the use of novel solid-source MOMBE approach paves the way to the atomic-layer controlled synthesis of complex oxides of "stubborn" metals, which are not only difficult to evaporate but also hard to oxidize.

# 1. INTRODUCTION

RuO$_2$ has gained considerable attention for the rich material properties it exhibits. High thermal conductivity, strong resistance to chemicals, and high electrical conductivity resulted in RuO$_2$ being used historically in thermometers[1], integrated circuits[2] and plasmonic[3], and as electrodes in supercapacitors.[4-5] RuO$_2$ also shows excellent catalytic properties, for example being a highly active oxygen evolution reaction catalyst.[6-7] When alloyed with metals like La, RuO$_2$ exhibits decrease in thermal conductivity and therefore has been argued to be potentially useful in thermoelectric devices.[8-9] More recently, several novel phenomena have also emerged in RuO$_2$ films, such as itinerant antiferromagnetism, and strain-stabilized superconductivity (figure 1).[10-11] Additionally, RuO$_2$ also serves as a precursor to the growth of more complex materials such as perovskite SrRuO$_3$ and Sr$_2$RuO$_4$, which are shown to exhibit itinerant ferromagnetism and unconventional superconductivity, respectively.

Bulk RuO$_2$ stabilizes in the prototypical tetragonal rutile crystal structure (figure 1) with space group #136, $P4_2/mnm$ and with lattice parameters $a = b = 4.492$ Å and $c = 3.106$ Å.[12] For thin films, RuO$_2$ has mostly been prepared as polycrystalline films[13-17] although, recently, several papers on epitaxial crystalline films have been reported.[10-11, 18-24] Molecular beam epitaxy (MBE) is one of these thin film techniques, which has been used for the growth of epitaxial single crystalline RuO$_2$[10-11, 18-20]. In general, the growth of Ru-based oxides in oxide MBE is challenging due to the ultra-low vapor pressure and low oxidation potential of Ru. Temperatures exceeding 2000°C are needed to achieve a suitable Ru vapor pressure for growth which is why reports of MBE-grown RuO$_2$ have all used electron-beam (e-beam) evaporators. Furthermore, the low oxidation potential makes stabilizing the Ru$^{4+}$ in RuO$_2$ difficult and has led to ozone, a highly oxidizing source, being the preferred oxidant in most reports. While the use of e-beam source and ozone has facilitated synthesizing epitaxial $RuO_2$ or other Ru-based oxides, they possess several challenges associated with the issues related to the instability of fluxes, and source oxidation in the presence of ozone.

In this paper, we demonstrate a novel solid source metal-organic molecular beam epitaxy (MOMBE) approach using a detailed $RuO_2$ growth study that addresses these synthesis issues by using a solid metal-organic precursor as the metal Ru source.[25] This allows for supplying a "pre-oxidized" metal with orders of magnitude higher vapor pressure than the elemental metal at a particular temperature.

## 2. EXPERIMENTAL SECTION

$RuO_2$ films were grown using the solid source MOMBE approach which has been described in more detail elsewhere.[25] Here, an effusion cell source temperature of 100°C was used for $Ru(acac)_3$ and oxygen was supplied with a radio-frequency inductively coupled plasma source. An oxygen background pressure of ~ $10^{-6} – 10^{-5}$ Torr was used. Substrate temperatures were 300°C, unless stated otherwise. Films were grown on *r*-plane sapphire *r*-$Al_2O_3$, $TiO_2$ (101), $TiO_2$ (110), $TiO_2$ (001), and $TiO_2$ (100) substrates.

Film surfaces were monitored before, during, and after growth using reflection high-energy electron diffraction (RHEED, Staib Instruments). Surface morphologies were imaged post-growth using atomic force microscopy (AFM, Bruker). Structural characterization and determination of film thickness was carried out using high-resolution X-ray diffraction (HRXRD), reciprocal space mapping (RSM), and grazing incidence X-ray reflectivity (GIXR, Rigaku SmartLab XE). Thickness was also alternatively determined by finite thickness fringes if present in the HRXRD scans. X-ray photoelectron spectroscopy (XPS, Physical Electronics VersaProbe III) was used for determine the Ru oxidation state. Four-probe resistivity measurements were performed in the Van der Pauw geometry as a function of temperature (Quantum Design DynaCool Physical Property Measurement System). Ohmic contacts were made using aluminum wire bonds.

Cross-sectional Scanning Transmission Electron Microscopy (STEM) samples were prepared by a Focused Ion Beam (FIB) method using FEI Helios Nanolab G4 dual-beam FIB, where lamellae were cut and thinned using 30 kV Ga-ion beam and further polished with 2 kV Ga-ion beam. Low-magnification high-angle annular dark-field STEM (HAADF-STEM) images were acquired using Thermofischer Talos F200X, and atomic-resolution HAADF-STEM images and Energy Dispersive X-ray (EDX) elemental maps were obtained using aberration-corrected FEI Titan G2 60-300 equipped with a super-X EDX detector. The microscopes were operated at 200 keV and screen current was ~25 pA. Probe semi convergent angle was 10.5 mrad and 17.3 mrad for Talos and Titan microscopes, and detector angles for HAADF-STEM images were in the range of 55-110 mrad.

## 3. RESULTS AND DISCUSSION

Figure 2a shows the vapor pressure of Ru metal in comparison to the precursor used for Ru in this study, Ru(acac)$_3$.[26] We also illustrate in figure 2b the important factors relevant to thin film synthesis by comparing them between e-beam-assisted MBE, hybrid MBE, and the solid-source MOMBE. Clearly, besides the advantages of high vapor pressure and the pre-oxidized state of the metal, the solid-source MOMBE also does not use a liquid precursor like hybrid MBE.[27] The solid metal-organic precursor can be sublimed in a conventional low temperature effusion cell directly in the vacuum system instead of a relatively complex gas inlet system. The low temperature effusion cell is also significantly less expensive, less complicated, and safer to operate than an e-beam source. The Ru(acac)$_3$ precursor itself comes with an additional source of oxygen bonded in the ligands, is air stable, and non-toxic, removing some safety concerns that can come with the use of metal-organic precursors like hexamethylditin (HMDT) in hybrid MBE growth of Sn-containing compounds.[28]

Using solid-source MOMBE approach, we first examined the effect of substrate temperature on the growth of $RuO_2$. $RuO_2$ films were grown on $r$-$Al_2O_3$ with substrate temperatures ($T_{sub}$) from 300°C to 850°C, for a fixed growth time. All films were epitaxial and phase pure with a single peak corresponding to $RuO_2$ (101) orientation, the common epitaxial orientation for rutile films on $r$-$Al_2O_3$ as shown in the HRXRD scans in figure 3a..[29] As $T_{sub}$ was increased, the growth rate increased, figure 3b, which led to differences in thickness of the films from 7 – 17 nm. As will be discussed it later, the films grown at $T_{sub}$ = 750°C and 850°C had surfaces too rough to determine a reliable thickness using GIXR. However, an estimate of the growth rates, and therefore thicknesses, is given in Figure 3b, obtained from the peak broadening of (101) film peak using the Scherrer formula.[30] While the Scherrer analysis can give a poor approximation of the film thickness, thicknesses obtained here agreed well with those from GIXR and HRXRD thickness fringes for T ≤ 650 °C.

The initial increase and later saturation of the growth rate with increasing $T_{sub}$ indicates a change of growth mechanism from a reaction-limited to a flux-limited regime.[31] This suggests that below 650°C the growth rate is limited by the thermal decomposition of the $Ru(acac)_3$ precursor. Above 650°C, the relatively constant growth rate is typical of being limited by the amount of precursor being supplied, or the flux.[31] No desorption-limited growth regime, i.e. a decrease in growth rate with increasing temperature, was observed. Nevertheless, the change from reaction- to flux-limited regime is not surprising and has been seen in other binary oxide systems grown by hybrid MBE approaches.[29, 32-33]

With increasing $T_{sub}$, an increase in the out-of-plane spacing of (101) planes ($d_{(101)}$) was seen reaching toward the expected bulk value. The change in $d_{(101)}$ with $T_{sub}$ is most likely due to the strain relaxation. To determine whether strain relaxation was due to the growth rate or film thickness, thicker films were grown at a constant growth rate, by keeping $T_{sub}$ = 300°C (figure S1). In this case, even as thickness was increased up to 48 nm, $d_{(101)}$ did not reach the bulk value. For instance, film thickness of 48 nm yielded $d_{(101)}$ = 2.538 ± 0.002 Å, which is significantly less than that of the 17 nm film ($d_{(101)}$ = 2.544 ± 0.002 Å) grown at higher temperatures (and at higher growth rate). This results thus suggest that the strain relaxation is more dominant effect at higher substrate temperature which is also accompanied by the higher growth rates.

Consistent with strain relaxation with increasing temperature, the full width at half maximum (FWHM) of the film (101) rocking curves increased by about an order of magnitude from 450°C to 550°C, as shown in figure 3d. RHEED images taken before growth, 10 minutes into growth, and after growth and cool down in oxygen ($T_{sub}$ = 200°C), as well as the post-growth AFM images, are shown in figures 3e – 3f. From the AFM images, it can be clearly seen that the increase in FHWM was also accompanied by a roughening of the film surface. The difference in the surface morphologies was confirmed by RHEED to be a result of a change in the growth mode during growth. At 10 minutes of growth, considerable differences in the RHEED patterns can be seen for these films grown at higher temperatures, with a change to an island growth mode. Irrespective of $T_{sub}$, XPS confirmed $Ru^{4+}$ valence states in these samples (figure S2).

Having identified the optimal substrate temperatures of 300°C – 450°C, RuO$_2$ films were grown at 300°C on TiO$_2$ substrates with different orientations. HRXRD scans, figure 4a, confirm phase pure, epitaxial, single crystalline films on all these substrates. Finite thickness fringes are present in all cases, although not very well defined in the case on TiO$_2$ (001), attesting to the high structural quality on a short lateral length scale. To investigate the structure of these films on an atomic scale, we performed STEM imaging of a representative $RuO_2$ film grown on $TiO_2$ (101) along [$\bar{1}$01] and [010] zone axes. Consistent with the HRXRD data, phase pure, epitaxial film is seen with a sharp film/substrate interface with no misfit dislocations. The lack of dislocations signifies coherent growth, which agrees well with the strained $d_{101}$ = 2.51 ± 0.002 Å obtained from HRXRD. EDX elemental maps further attest to a uniform distribution of Ru in the film.

Interestingly, STEM images also reveal an atomically smooth surface along the [010] zone axis whereas a significantly rougher morphology was observed when viewed along [$\bar{1}$01] zone axis (figure 4b). As shown in the zoom-in image of figure 4b, the rough surface was found to be terminated not only at the expected (101) plane parallel to the (101) $TiO_2$ substrate but also other plane consistent with (111) face. While the origin of this unusual surface morphology is unclear, and remains a subject of future study, we argue that it may be related to the significantly different strain mismatch of +0.04% and +2.3%, along [$\bar{1}$01] and [010] direction, respectively.

As a next step, we investigated the strain relaxation behavior of $t$ nm $RuO_2$ film/$TiO_2$ (110) with t = 3 - 26 nm. Theoretically, $RuO_2$ on $TiO_2$ (110) has a relatively large lattice mismatch of about - 4.7% and + 2.3% along the [001] and [1$\bar{1}$0] directions, respectively, indicating coherently strained growth may be challenging on $TiO_2$ (110). Figure 5a shows HRXRD scans of $t$ nm $RuO_2$ film/$TiO_2$ (110) with t = 3 - 15 nm revealing thickness fringes and a film (110) peaks being partially overlapped with that of the substrate. The well-defined Kiessig fringes again attest to the high-quality film on a short lateral length scale. Upon analysis of the film peak position, 26 nm $RuO_2$ film/$TiO_2$ (110) yielded $d_{(110)}$ = 3.204 ± 0.002 Å, which is larger than the bulk value of 3.176 Å suggesting partially strained films. To examine the strain state of these films along in-plane [001] and [1$\bar{1}$0] directions, RSMs were taken. Figures 5(b-e) show RSMs around (332) and (310) reflections for two representative films with $t$ = 6 nm and 26 nm. As was expected based on the value of $d_{(110)}$, the 26 nm sample was partially relaxed with both in-plane spacings, along the [001] and [1$\bar{1}$0] directions, falling between the expected fully strained and fully relaxed values (figures 5(b-d)). Interestingly, the 6 nm sample showed the same in-plane spacing as the 26 nm sample along the [001] direction (the [1$\bar{1}$0] direction could not be determined due to overlap with the substrate). These results suggest the strain relaxation begins to occur at $t$ as small as 6 nm or less for film grown on TiO$_2$ (110). Consistent with the strain-relaxation behavior, a broadening of the (220) RuO$_2$ film rocking curve was seen with increasing $t$. Figures 5f and 5g show the rocking curves of 26 nm and 6 nm film, respectively. Rocking curves were fitted using two Gaussian peaks, marked as a broad and a narrow peak. The results of this fitting are shown in figure 5h and 5i. Figure 5h shows the narrow peak remained at a relatively constant FWHM of ~ 0.07° while the broad peak FWHM decreased. Taking the ratio of the peak intensity of the broad component ($I_{broad}$) to the total intensity of the two ($I_{total} = I_{broad} + I_{narrow}$), figure 5i, revealed an almost linear increase in this ratio with increasing film thickness. Results from the identical analysis of $t$ nm RuO$_2$ film/$r$-Al$_2$O$_3$ are also included in figure 5i, which compare well with films grown on TiO$_2$ (110) substrates.

The origin of the broad component can be thought of as being caused by the disorder induced due to the strain relaxation. We know the strain relaxation process has begun by at least 6 nm based on the RSM results, and likely at an even smaller thickness because of the presence of the broad component in the rocking curves of those films as well. As thickness is increased, the volume fraction of film that was influenced by the relaxation increases and, therefore, the intensity of the broad component does as well. Similar results were seen in films grown on $r$-Al$_2$O$_3$, however, with a faster increase in the intensity ratio with increasing $t$. This observation is again consistent with faster strain relaxation expected from a larger lattice-mismatch and difference in symmetry between $RuO_2$ and *r*-$Al_2O_3$.

Finally, we turn to the discussion of electrical property of these films revealing a clear correlation between film thickness, strain relaxation and electrical resistivity. Figure 6a shows the temperature-dependent resistivity ($\rho$) for *t* nm $RuO_2$ film/$TiO_2$ (110) with t = 3 - 15 nm. Films with *t* > 15 nm showed a large resistance anisotropy between the two in-plane directions for reliable four-terminal resistivity measurements, which is consistent with the prior results[10]. All films showed metallic behavior with increasing resistivity with decreasing *t*. The room-temperature $\rho$ of 15 nm was 56 μΩ·cm, closest to the 35 μΩ·cm of bulk $RuO_2$ among the films grown on $TiO_2$ (110)[34]. Figure 6b shows the residual resistivities ($\rho_0 = \rho$ at 1.8 K) revealing an exponential-like decrease with increasing *t*, saturating at ~ 13 μΩ·cm. Increasing film thickness can influence the electronic properties by way of finite size effects, such as effects from the film-substrate interface, dimensionality as well as the defects arising from the strain relaxation process, such as misfit and/or threading dislocations. With regards to the latter, as strain relaxation occurs and more defects are formed, the residual resistivity should likely increase as it is generally dependent on these structural crystalline defects. Here we see the opposite trend, implying the increase in thickness has a much larger effect on $\rho$ than strain relaxation-related defect formation. To this end, we plot $\rho_0$ vs. the film rocking curve intensity ratio we defined earlier in Figure 6c. This shows, once again, the opposite trend of what should be seen if strain relaxation was the critical factor. These data establish an important role of film dimensionality on the electrical transport properties of $RuO_2$ films.

## 4. CONCLUSIONS

In summary, we have shown the growth of epitaxial RuO$_2$ films of different orientations using different substrates by a novel solid-source MOMBE. Single crystalline films with low film rocking curve FWHMs were grown on *r*-Al$_2$O$_3$ with substrate temperatures between 300°C - 450°C. At higher temperatures, films showed a significant increase in structural disorder. Using this approach and by keeping substrate temperature of 300°C, films were then grown on TiO$_2$ substrates with different orientations. STEM results confirmed phase pure, epitaxial films free of strain-relaxation-related defects when grown on TiO$_2$ (101). However, films on TiO$_2$ (110) were found to relax at thickness as low as 6 nm. Films with resistivity similar to that of the bulk RuO$_2$ single crystals was obtained for 15 nm RuO$_2$/TiO$_2$ (110). Finally, with increasing film thickness, we revealed an important role of film thickness on electrical properties. This work establishes the solid-source MOMBE technique for the growth of high-quality RuO$_2$ in a much simpler, and cost-effective manner when compared to conventional MBE approaches. For instance, Ru metal was supplied by subliming a solid metal-organic precursor, Ru(acac)$_3$ in a low temperature effusion cell operating at 100°C as opposed to several thousand using e-beam in conventional MBE.

**Acknowledgments**

This work was primarily supported by the U.S. Department of Energy through DE-SC002021. The work also benefitted from the Norwegian Centennial Chair Program (NOCC) and Vannevar Bush Faculty Fellowship. S.N. and A.K.M. acknowledge support from the Air Force Office of Scientific Research (AFOSR) through Grant Nos. FA9550-19-1-0245 and FA9550-21-1-0025 and partially through NSF DMR-1741801. D.L. acknowledge support from the U.S. DOE through the University of Minnesota Center for Quantum Materials, under Award No. DE-SC0016371. A.R., H.Y. and K.A.M. acknowledge support from the UMN MRSEC program under award no. DMR-2011401. Parts of this work were carried out at the Characterization Facility, University of Minnesota, which receives partial support from NSF through the MRSEC program under award no. DMR- 2011401.

**Author contributions:** W.N. and S.N. grew samples and performed HRXRD and AFM characterization. A.K.M., and A.R performed temperature dependent transport measurements. D.L performed XPS measurements. H.Y. performed TEM sample preparation and STEM imaging under the guidance of K.A.M. W.N. and B.J. wrote the manuscript. All authors contributed to the discussion and manuscript preparation.

**Competing interests:** The authors declare no competing interests.

**Data and materials availability:** All data needed to evaluate the conclusions of the paper are present in the paper and/or the Supplementary Materials. Additional data related to this paper may be requested from the authors.

**Supplementary Information:** See supplementary material at [URL] for HRXRD of $RuO_2$ films as a function of thickness, for the XPS revealing valence state of Ru.

**Figures (Color Online)**

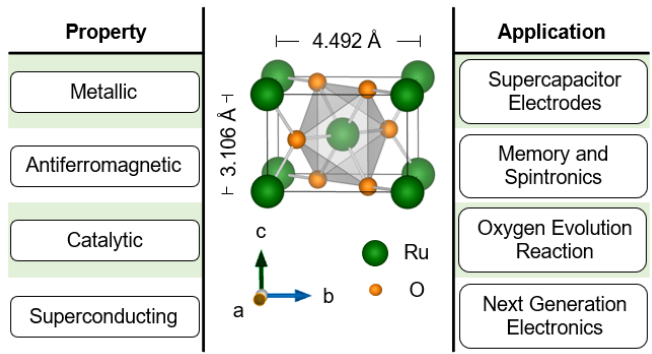

**Figure 1.** Rutile crystal structure and lattice parameters for RuO$_2$, a simple binary oxide, which exhibits many different property-application relationships.

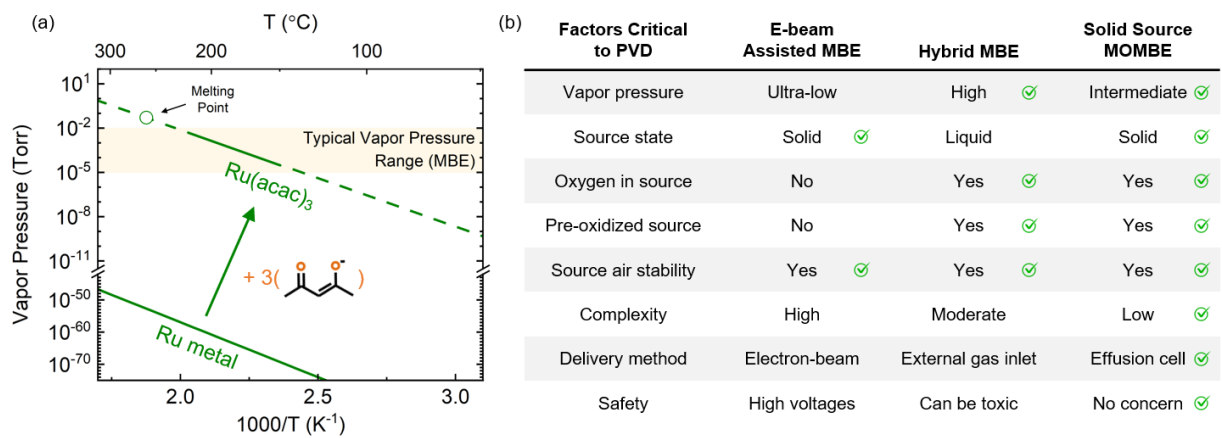

**Figure 2.** (a) Vapor pressure of Ru metal and the metal-organic precursor, Ru(acac)$_3$. (b) Table of a few of the factors which are critical to the physical vapor deposition approaches, compared for electron-beam assisted MBE, hybrid MBE, and solid-source MOMBE. Green symbol signifies desirable.

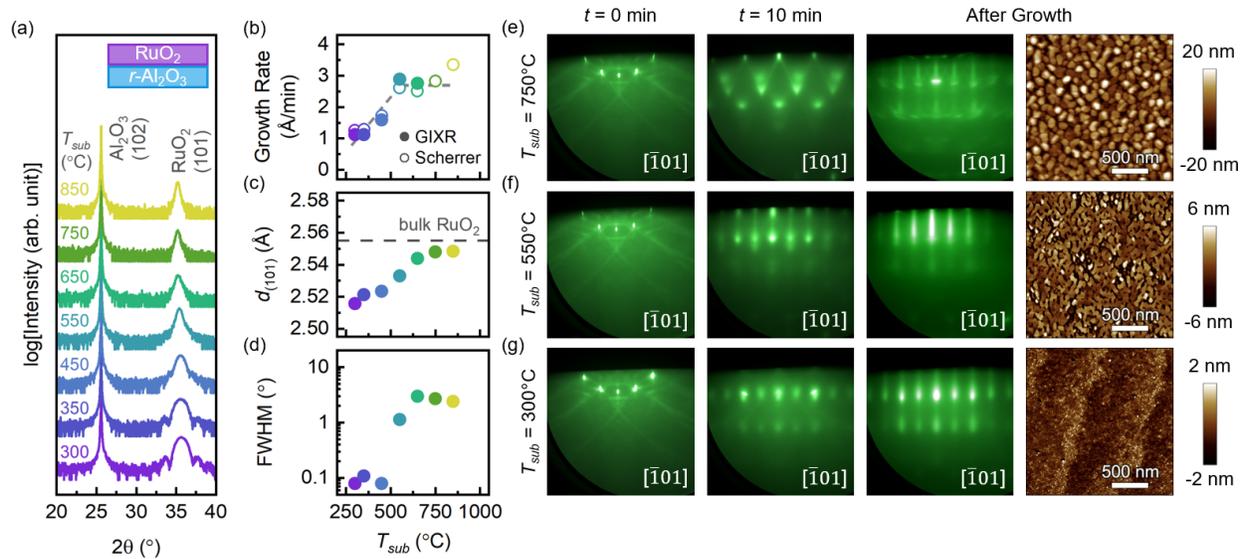

**Figure 3.** (a) HRXRD patterns for $RuO_2$ (101) films grown on $r$-$Al_2O_3$ with increasing substrate temperature from bottom to top. (b) Growth rate, (c) out-of-plane (101) plane spacing, $d_{(101)}$ and (d) FWHM of the film (101) rocking curve peak. RHEED along the film $[\bar{1}01]$ azimuth and AFM images before, 10 minutes into, and after growth for substrate temperature of (e) 750°C, (f) 550°C, and (g) 300°C.

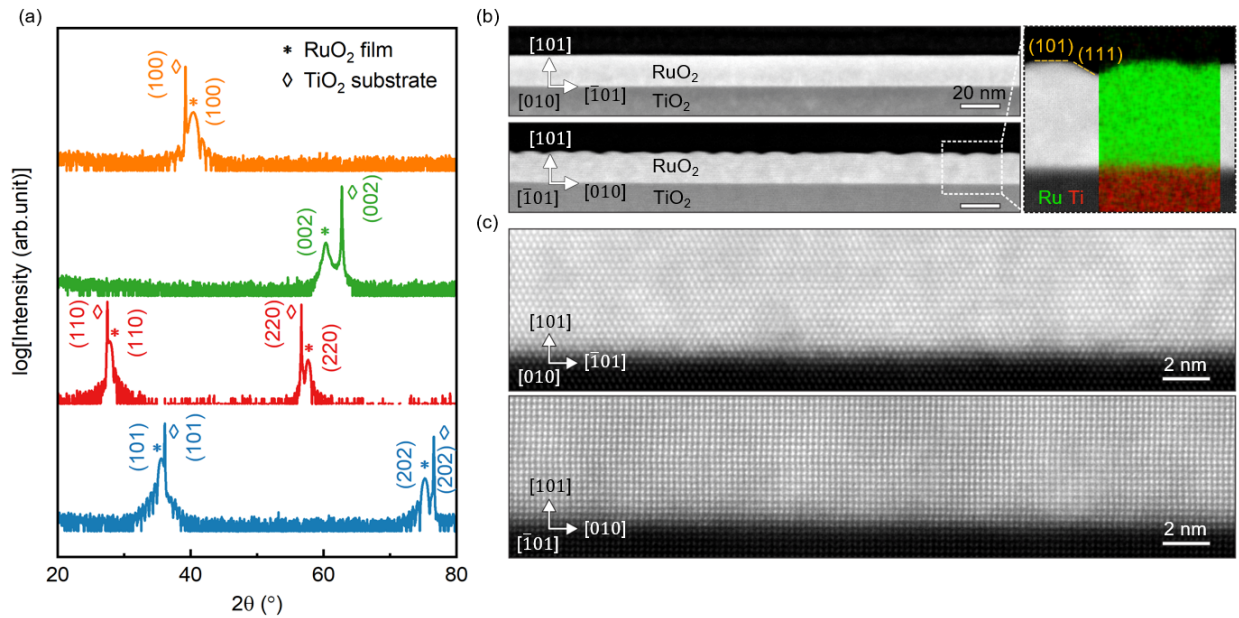

**Figure 4.** (a) HRXRD patterns of RuO$_2$ films grown on a variety of TiO$_2$ substrate orientations, (101), (110), (001), and (100). Film thicknesses are 16 nm, 12 nm, 20 nm, and 10 nm, from bottom to top. (b) Cross-sectional HAADF-STEM images of 16 nm RuO$_2$ thin films grown on TiO$_2$ (101) in the [010] and [$\bar{1}$01] directions and STEM-EDX elemental map. The EDX map was constructed using Ru $L_\alpha$ and Ti $K_\alpha$ emissions. (c) Atomic-resolution HAADF-STEM images of the RuO$_2$/TiO$_2$ interface along the two different crystallographic orientations.

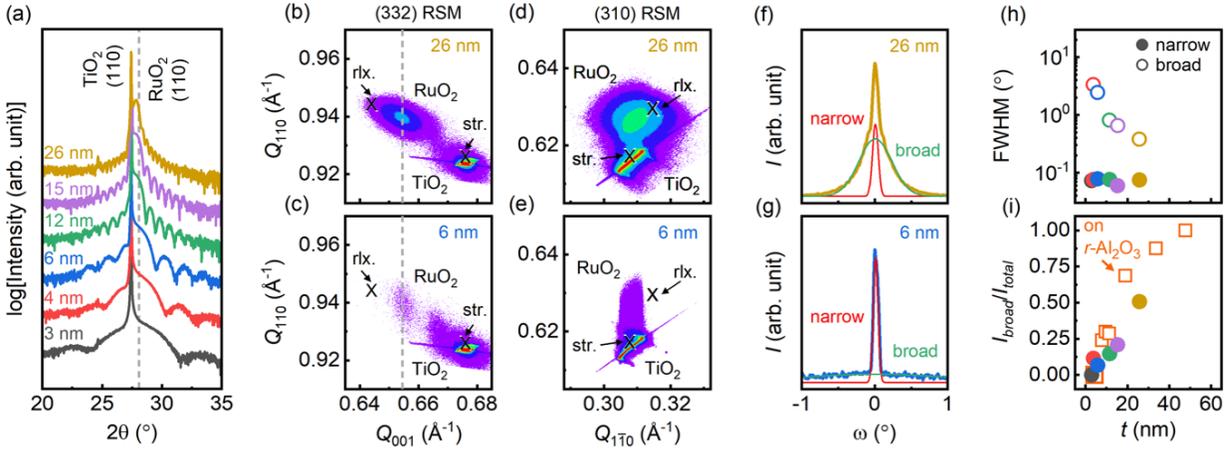

**Figure 5.** (a) HRXRD patterns of RuO$_2$ films grown on TiO$_2$ (110) substrates with increasing thickness, from bottom to top. (332) RSM for (b) 26 nm and (c) 6 nm film. (310) RSM for (d) 26 nm and (e) 6 nm film. Film (220) rocking curves for (f) 26 nm and (g) 6 nm film. Gaussian fits are shown for the "narrow" and "broad" peaks. In RSMs, str. and rlx. correspond to a fully-strained and a fully-relaxed position, respectively. Thickness-dependent (h) FWHM of the narrow and broad fits and (i) intensity ratio of broad peak intensity to total intensity for (220) film peaks.

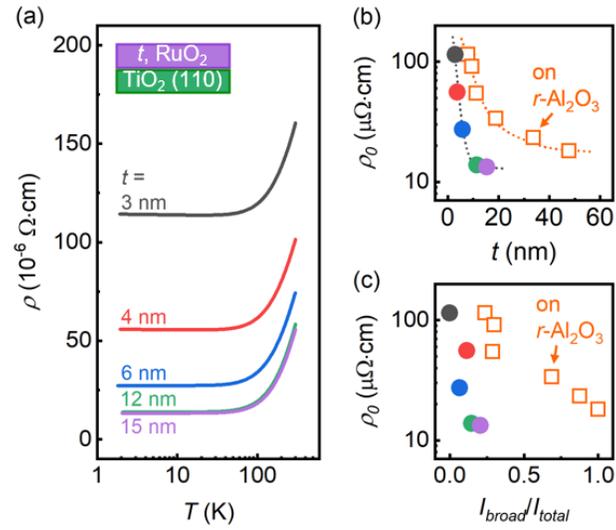

**Figure 6.** (a) Resistivity (ρ) vs. temperature for RuO$_2$ films grown on TiO$_2$ (110) as a function of film thickness. Residual resistivity ($\rho_0$), taken at 1.8 K, for RuO$_2$ grown on TiO$_2$ (110) (filled circles) and on *r*-Al$_2$O$_3$ (open squares) as a function of (b) film thickness and (c) intensity ratio of broad peak intensity to total intensity for (220) film peaks.